\documentclass[12pt,a4paper]{article}

\title{Connecting connections\\
{\small A bricklayer view of General relativity}}

\author{Llu\'{\i}s\ Bel\\
\emph{wtpbedil@lg.ehu.es}
}
\begin{document}

\maketitle

\begin{abstract}

As true as it is that a bricklayer needs a plumb line and a T-square, so it is that a physicist using general relativity needs how to draw geodesics and use fields of congruent vector frames of reference. While the first part of the preceding statement depends on the Christoffel connection and related metric and curvature concepts, the second part depends on the Weitzenb\"{o}ck connection and the concept of torsion. This dual structure has been considered before as a possibility of using either one of them to describe General relativity. We claim here that both structures have to be correlated to produce useful interpretations of any space-time model.

\end{abstract}

\section*{Introduction}

Weitzenb\"{o}ck is known for disliking Frenchmen in general and E. Cartan in particular\,\footnote{Read about a  Weitzenb\"{o}c's acrostic in Sect. 13  of \cite{Schucking1}} but he is also known for his achievements as a mathematician. In fact both men have contributed to improve our understanding of the geometrical formalism of General relativity. While Cartan generalized Riemannian geometry considering it as a particular case of a geometry with a linear connection, Weitzenb\"{o}ck introduced a new connection in the formalism which is only indirectly related to space-time intervals and emphasizes the importance of the concept of torsion instead of the concept of curvature.

While the usefulness of the curvature tensor is rather well-understood in physics, the meaning and the usefulness of the torsion tensor is less clear. A frequent point of view assumes that torsion is a manifestation of the spin of the particles making up the sources of the gravitational field. From this point of view if there is no spin we can altogether forget about torsion.

Another point of view presented in \cite{Pereira1} and \cite{Pereira2} suggests that while the formalisms based on the Christoffel connection or the Weitzenb\"{o}ck connection are strictly equivalent from a mathematical point of view they are inequivalent from a physical point of view because only the latter can be understood as the formalism of a gauge theory, and while the concept of force does not exist when using the first formalism the second one reestablishes it. Still more recently it has been claimed in \cite{Schucking1} and \cite{Schucking2} that gravitation is not curvature but torsion.

In this paper we do not try to decide who was first or who is more palatable: the hen or the egg? and claim that there exists something like a hen-egg concept. This meaning that the two  connections, Christoffel's and Weitzenb\"{o}ck's, do not have to be considered as options of an alternative, but that in the contrary they have to be correlated and used jointly. There is no new physics in this process but doing it appropriately we hope to improve our knowledge of General relativity.

The first and second sections are mainly a remainder of known definitions and general results. In the third section we define Special  Weitzenb\"{o}ck connections for which the contortion and the torsion tensors are equal, and we introduce also the concept of Restricted covariance. In the fourth section we introduce the generalization of a tensor that we introduced in a preceding paper \cite{Bel} in the framework of the linear approximation where it proved to have some relevance in defining the energy-momentum of gravitational fields. The fifth section is an application of the general results reminded or derived in the preceding ones to the theory of the linear approximation in General relativity. This provides our paper \cite{Bel} with a well-founded mathematical formalism. 

The epilogue deals with Weitzenb\"{o}ck-like connections defined on 3-di\-mensional  quotient manifolds and contains a proposal to generalize the concept of rigid motions defined by Born.

 
\section{Generalities about Linear connections}

{\bf General Cartan connections}\,\footnote{Essentially drawn from Chap. IV of \cite{Lichnerowicz}}
\vspace{0.5cm}

Let $V_n$ be a differential manifold of dimension $n$ and let $D_n$ and $D^\prime_n$ be two intersecting domains of $V_n$, with local coordinates $x^\alpha$ and $x^{\alpha^\prime}$, with greek indices running from $1$ to $n$. A Linear connection on $V_n$ is a field of 1-forms of type (1,1):  

\begin{equation}
\label{0.1}
\omega^\alpha_\beta(x)=\Gamma^\alpha_{\beta\gamma}(x)dx^\gamma
\end{equation}
such that at any point of the intersection of any two domains one has:

\begin{equation}
\label{0.2}
\omega^{\alpha^\prime}_{\beta^\prime}(x^\prime)=A^{\alpha^\prime}_\rho(x(x^\prime))\omega^\rho_\sigma(x(x^\prime))
A^\sigma_{\beta^\prime}(x^\prime)+A^{\alpha^\prime}_\rho(x(x^\prime))dA^\rho_{\beta^\prime}(x^\prime)
\end{equation}
with:

\begin{equation}
\label{0.3}
A^{\alpha^\prime}_\rho(x(x^\prime))=\frac{\partial x^{\alpha^\prime}}{\partial x^{\rho}}(x(x^\prime)), \quad
A^\sigma_{\beta^\prime}(x^\prime)=\frac{\partial x^\sigma}{\partial x^{\beta^\prime}}(x^\prime)
\end{equation}
where:

\begin{equation}
\label{1.3}
x^{\rho^{\prime}}=x^{\rho^{\prime}}(x),\ x^\sigma=x^\sigma(x^\prime)
\end{equation}
define the coordinate transformation.

The covariant derivatives of a function $f$ and  a covariant vector vector field $v_\alpha$ are by definition:

\begin{equation}
\label{0.3.1}
\nabla_\beta f=\partial_\beta f, \quad \nabla_\beta v_\alpha=\partial_\beta v_\alpha-\Gamma^\rho_{\alpha\beta}v_\rho
\end{equation}
from where, using Leibniz rule to calculate derivatives of tensor products, follows the general formula to define the covariant derivative of any tensor.

The Curvature of the linear connection $\omega^\alpha_\beta$ is the 2-form of type(1,1):

\begin{equation}
\label{0.4}
\Omega^\alpha_\beta=\frac12 R^\alpha_{\beta\gamma\delta}dx^\gamma\wedge dx^\delta
\end{equation}
where:

\begin{equation}
\label{0.5}
R^\alpha_{\beta\gamma\delta}=
\partial_\gamma\Gamma^\alpha_{\beta\delta}-\partial_\delta\Gamma^\alpha_{\beta\gamma}+ 
\Gamma^\alpha_{\rho\gamma}\Gamma^\rho_{\beta\delta}-\Gamma^\alpha_{\rho\delta}\Gamma^\rho_{\beta\gamma}
\end{equation}
and the Torsion is the vector-valued 2-form:

\begin{equation}
\label{0.6}
\Sigma^\alpha=\frac12 T^\alpha_{\beta\gamma}dx^\beta\wedge dx^\gamma
\end{equation}
where:

\begin{equation}\label{0.7}
T^\alpha_{\beta\gamma}=
-(\Gamma^\alpha_{\beta\gamma}-\Gamma^\alpha_{\gamma\beta})
\end{equation}

From these definitions and from the fact that $d^2=0$, $d$ being the exterior differentiation operator, there follow the following identities:

\begin{equation}
\label{0.8}
\nabla_{[\epsilon} R^\alpha_{\beta|\gamma\delta]}=R^\alpha_{\beta\rho[\epsilon}T^\rho_{\gamma\delta]} 
\end{equation}
where $[.| |..]$ is here the three indices complete anti-symmetrization operator. And:

\begin{equation}
\label{0.9}
{\nabla}_{[\alpha}{T}^\rho_{\beta\gamma]} 
+{T}^\rho_{\sigma[\alpha}{T}^\sigma_{\beta\gamma]}=R^\rho_{[\alpha\beta\gamma]}
\end{equation}

The last general definition that we shall need to mention is that of an auto-parallel, which is any parameterized curve $x^\alpha(\lambda)$ solution of the differential equations:

\begin{equation}
\label{0.10}
\frac{d^2x^\alpha}{d\lambda^2}+\Gamma^\alpha_{\beta\gamma}\frac{dx^\beta}{d\lambda}\frac{dx^\gamma}{d\lambda}=a\frac{dx^\alpha}{d\lambda}
\end{equation}
where the function a in the r-h-s  depends on the parameter $\lambda$. If it is zero the parameter is said to be an affine parameter.
\vspace{0.5cm}


{\bf Weitzenb\"{o}ck connections\,\footnote{Mainly drawn from \cite{Pereira1} and \cite{Pereira2}}}

\vspace{0.5cm}
With latin indices running also from $1$ to $n$, let $\theta^a_\alpha$ be $n$ linearly independent 1-forms, $\theta^\beta_b$ being its duals contravariant vector fields: 

\begin{equation}
\label{1.1}
\theta^a_\alpha\theta^\alpha_b=\delta^a_b\Leftrightarrow \theta^a_\beta\theta^\alpha_a=\delta^\alpha_\beta
\end{equation}

A Weitzenb\"{o}ck connection $\widetilde\Gamma^\lambda_{\beta\gamma}$ associated with with $\theta^a_\alpha$ is defined as the connection that leads to the covariant derivative with symbol $\widetilde\nabla$:

\begin{equation}
\label{1.1.0}
\widetilde\nabla_\alpha \theta_\beta^a=\partial_\alpha \theta_\beta^a-\widetilde\Gamma^\rho_{\beta\alpha}\theta_\rho^a=0
\end{equation}
and therefore we have :

\begin{equation}
\label{1.12}
\widetilde\Gamma^\lambda_{\beta\gamma}=\theta^\lambda_a\partial_\gamma\theta^a_\beta
\end{equation}

The most significant property of Weitzenb\"{o}ck connections is that their curvature tensor is zero:

\begin{equation}
\label{1.22}
\widetilde R^\alpha_{\beta\gamma\delta}=
\partial_\gamma\widetilde\Gamma^\alpha_{\beta\delta}-\partial_\delta\widetilde\Gamma^\alpha_{\beta\gamma}+ 
\widetilde\Gamma^\alpha_{\rho\gamma}\widetilde\Gamma^\rho_{\beta\delta}-\widetilde\Gamma^\alpha_{\rho\delta}\widetilde\Gamma^\rho_{\beta\gamma}=0
\end{equation} 

On the other hand its torsion is:

\begin{equation}
\label{1.15}
T^\lambda_{\beta\gamma}=-(\widetilde\Gamma^\lambda_{\beta\gamma}-\widetilde\Gamma^\lambda_{\gamma\beta})=
\theta^\lambda_a(\partial_\beta\theta^a_\gamma-\partial_\gamma\theta^a_\beta)
\end{equation}
And taking into account (\ref{1.22}) the identities (\ref{0.9}) become:

\begin{equation}
\label{1.25}
{\widetilde\nabla}_{[\alpha}{T}^\rho_{\beta\gamma]} 
+{T}^\rho_{\sigma[\alpha}{T}^\sigma_{\beta\gamma]}=0
\end{equation}

The auto-parallels, referred to an affine parameter, are now the solutions of: 

\begin{equation}
\label{1.25.1}
\frac{d^2x^\alpha}{d\lambda^2}+\widetilde\Gamma^\alpha_{\beta\gamma}\frac{dx^\beta}{d\lambda}\frac{dx^\gamma}{d\lambda}=0
\end{equation}
and should be better called loxodromies.

\vspace{0.5cm}

{\bf Christoffel connections}

\vspace{0.5cm}
Let $g_{\alpha\beta}(x)$ be a Riemannian metric of any signature defined on $V_n$. The Christoffel connection associated with this metric is by definition that connection which has the following two properties:

\begin{equation}
\label{1.8.0}
\nabla_\gamma g_{\alpha\beta}=0, \quad \bar T^\rho_{\alpha\beta}=
\Gamma^\rho_{\beta\alpha}-\Gamma^\rho_{\alpha\beta}=0
\end{equation}
This leads to an unique symmetric connection with symbols:
 
\begin{equation}
\label{1.8}
\Gamma^\lambda_{\beta\gamma}=\Gamma_{\beta\gamma\alpha}g^{\lambda\alpha}
\end{equation}
with:
\begin{equation}
\label{1.9}
\Gamma_{\beta\gamma\alpha}=\frac12 (\partial_\beta g_{\gamma\alpha}
+\partial_\gamma g_{\beta\alpha}-\partial_\alpha g_{\beta\gamma})
\end{equation}

The Riemann tensor is the curvature tensor defined in (\ref{0.5}) with the corresponding general connection being substituted by the Christoffel connection above. If it is zero then the metric $g_{\alpha\beta}$ can be reduced by a coordinate transformation to a matrix of constants $\eta_{\alpha\beta}$. 

With the same substitution we obtain the auto-parallels of a Christoffel connection that are better called geodesics of the metric:

\begin{equation}
\label{1.21.1}
\frac{d^2x^\alpha}{d\tau^2}+\Gamma^\alpha_{\mu\nu}\frac{dx^\mu}{d\tau}\frac{dx^\nu}{d\tau}=0
\end{equation}
In this case affine parameters are proportional to the proper length of the curve when this length is not zero.
\vspace{0.5cm}


\section{Connecting connections}

Let us consider any non singular diagonal matrix whose elements $\eta_{ab}$ are $1$ or $-1$, the corresponding quadratic form having arbitrary signature.

To such matrix and any field of $n$ 1-forms $\theta^a_\alpha$, as we considered before, it can be associated a Riemannian metric:

\begin{equation}
\label{1.17.0}
g_{\alpha\beta}= \eta_{ab}\theta^a_\alpha\theta^b_\beta. 
\end{equation}
that has the same signature as $\eta_{ab}$. When referring to this formula we shall say that $\theta^a_\alpha$ is an orthonormal decomposition of $g_{\alpha\beta}$ and, the other way around, we shall say $g_{\alpha\beta}$ is the metric derived from  $\theta^a_\alpha$. Equivalent orthonormal decompositions are related by point dependent frame transformations:

\begin{equation}
\label{1.17.0.1}
{\theta^\prime}^a_\alpha(x)=R^a_b(x)\theta^b_\alpha(x), 
\end{equation}
such that:

\begin{equation}
\label{1.17.0.2}
R^a_c(x)R^b_d(x)\eta_{ab}=\eta_{cd}, 
\end{equation} 

Let us consider the two connections: the Weitzenb\"{o}ck connection $\widetilde\Gamma^\alpha_{\beta\gamma}$ associated with $\theta^b_\alpha(x)$ and the Christoffel connection $\Gamma^\alpha_{\beta\gamma}$ associated to the Riemannian metric defined in (\ref{1.17.0}). We have:

\begin{equation}
\label{1.17.0.3}
\nabla_\gamma g_{\alpha\beta}=0, \  \widetilde\nabla_\gamma g_{\alpha\beta}=0
\end{equation}
The first comes from the definition (\ref{1.8.0}) and using (\ref{1.17.0}) a short calculation proves the second. They share also a second property, namely:

\begin{equation}
\label{1.17.0.4}
\nabla_\rho\eta_{\alpha_1\cdots\alpha_n}=0, \  \widetilde\nabla_\rho\eta_{\alpha_1\cdots\alpha_n}=0
\end{equation}
where $\eta_{\alpha_1\cdots\alpha_n}$ is the volume element associated with the Riemannian metric (\ref{1.17.0})

Instrumental in connecting, so to speak, the two connections is the definition of the so-called Contortion, that is the tensor:

\begin{equation}
\label{1.17}
K^\lambda_{\beta\gamma}=\widetilde\Gamma^\lambda_{\beta\gamma}-\Gamma^\lambda_{\beta\gamma}
\end{equation}

Subtracting the two equations (\ref{1.17.0.3}) we obtain:

\begin{equation}
\label{1.17.1}
\nabla_\gamma g_{\alpha\beta}-\widetilde\nabla_\gamma g_{\alpha\beta}=
K^\rho_{\alpha\gamma} g_{\rho\beta}+K^\rho_{\beta\gamma} g_{\rho\alpha}=0
\end{equation}
and therefore defining:

\begin{equation}
\label{1.17.2}
K_{\alpha\beta\gamma}=K^\rho_{\beta\gamma} g_{\rho\alpha}, \quad
\end{equation}
we have:

\begin{equation}
\label{1.21}
K_{\alpha\beta\gamma}=-K_{\beta\alpha\gamma}
\end{equation}
i.e.: the fully covariant contortion is skew-symmetric with respect to the first pair of indices.

Defining similarly:

\begin{equation}
\label{1.17.2.1}
T_{\beta\gamma\alpha}=T^\rho_{\beta\gamma}g_{\rho\alpha}, 
\end{equation}
from Eqs. (\ref{1.8.0}), (\ref{1.15}) and (\ref{1.17}) we obtain:

\begin{equation}
\label{1.23}
K_{\alpha\beta\gamma}-K_{\alpha\gamma\beta}=-T_{\beta\gamma\alpha}
\end{equation}
from where it follows, using the Christoffel (+,+,$-$) algorithm:

\begin{equation}
\label{1.24}
K_{\alpha\beta\gamma}=\frac12(T_{\alpha\beta\gamma}-T_{\beta\gamma\alpha}-T_{\gamma\alpha\beta})
\end{equation}

Using (\ref{1.22}), (\ref{0.5}), (\ref{1.17}) and formulas like:

\begin{equation}
\label{1.24.1}
\widetilde\nabla_\rho K^\alpha_{\beta\gamma}=\partial_\rho K^\alpha_{\beta\gamma}
+\widetilde\Gamma^\alpha_{\sigma\rho}K^\sigma_{\beta\gamma}-\widetilde\Gamma^\sigma_{\beta\rho}K^\alpha_{\sigma\gamma}
-\widetilde\Gamma^\sigma_{\gamma\rho}K^\alpha_{\beta\sigma}
\end{equation}
a short calculation proves that:

\begin{equation}
\label{1.26}
R^\alpha_{\beta\gamma\delta}=-{\widetilde\nabla}_\gamma K^\alpha_{\beta\delta}+{\widetilde\nabla}_\delta K^\alpha_{\beta\gamma}
-K^\rho_{\beta\gamma}K^\alpha_{\rho\delta}+K^\rho_{\beta\delta}K^\alpha_{\rho\gamma}
-T^\rho_{\gamma\delta}K^\alpha_{\beta\rho}
\end{equation}
and then:

\begin{equation}
\label{1.28}
R_{\beta\delta}=-{\widetilde\nabla}_\alpha K^\alpha _{\beta\delta}+{\widetilde\nabla}_\delta K_\beta
-K^\rho_{\beta\sigma}K^\sigma _{\rho\delta}+K^\rho_{\beta\delta}K _\rho
-T^\rho_{\sigma  \delta}K^\sigma _{\beta\rho}
\end{equation}
and:
\begin{equation}
\label{1.29}
R=2\widetilde\nabla_\alpha K^\alpha-K^\rho K_\rho
-(K^\rho_{\mu\sigma}K^\sigma_{\rho\nu}+T^\rho_{\sigma\nu}K^\sigma_{\mu\rho})g^{\mu\nu}
\end{equation}
where $R_{\alpha\beta}$ is the Ricci tensor of the Riemannian metric; $R$ is the curvature scalar and where we have defined:

\begin{equation}
\label{1.29.1}
K_\beta=K^\alpha_{\beta\alpha}, \ K^\alpha=-K^\alpha_{\beta\gamma}g^{\beta\gamma}
\end{equation}

Moreover contracting the indices $\gamma$ and $\alpha$ in (\ref{1.26}) and using  (\ref{1.28}) and (\ref{1.29}) we obtain the final formula that we shall use later on:

\begin{equation}
\label{1.30}
\widetilde\nabla_\alpha H^\alpha_{\beta\delta}=Y_{\beta\delta}-S_{\beta\delta}
\end{equation}
where:

\begin{equation}
\label{1.30.1}
H^\alpha_{\beta\delta}=K^\alpha_{\beta\delta}-\delta^\alpha_\delta K_\beta
+g_{\beta\delta} K^\alpha,
\end{equation}
$S_{\beta\delta}$ being the Einstein tensor, and $Y_{\beta\delta}$ is a shorthand for:

\begin{equation}
\label{1.31}
Y_{\beta\delta}= 
-K^\sigma_{\beta\rho}K^\rho_{\delta\sigma}+K^\rho_{\beta\delta}K_\rho+\frac12 g_{\beta\delta}(K^\rho K_\rho
+K^\rho_{\nu\sigma}K^\sigma_{\mu\rho}g^{\mu\nu})
\end{equation}
To get this simplified formula we used (\ref{1.23}).

Possibly interesting is also the fact that using (\ref{1.17}) the geodesic equations (\ref{1.21.1}) of the Riemannian metric (\ref{1.17.0}) can be written, using an affine parameter:

\begin{equation}
\label{1.21.2}
\frac{d^2x^\alpha}{d\tau^2}+\widetilde\Gamma^\alpha_{\mu\nu}\frac{dx^\mu}{d\tau}\frac{dx^\nu}{d\tau}=
K^\alpha_{\mu\nu}\frac{dx^\mu}{d\tau}\frac{dx^\nu}{d\tau}
\end{equation}
and the auto-parallels equations (\ref{0.10}) of the Weitzenb\"{o}ck connection (\ref{1.12}) can be written:

\begin{equation}
\label{1.21.3}
\frac{d^2x^\alpha}{d\lambda^2}+\Gamma^\alpha_{\mu\nu}\frac{dx^\mu}{d\lambda}\frac{dx^\nu}{d\lambda}=
-K^\alpha_{\mu\nu}\frac{dx^\mu}{d\lambda}\frac{dx^\nu}{d\lambda}
\end{equation}


\section{Special Weitzenb\"{o}ck connections}

We shall say that the Weitzenb\"{o}ck connection is Special if:

\begin{equation}
\label{3.2}
\eta_{ab}\theta^a\wedge d\theta^b=0
\end{equation}
that includes the integrable case:

\begin{equation}
\label{3.2.1}
\theta^a\wedge d\theta^a=0,\ a=1\cdots n
\end{equation}

From:

\begin{equation}
\label{1.31.1}
T_{\alpha\beta\gamma}=g_{\gamma\rho}\theta^\rho_a(\partial_\alpha\theta^a_\beta-\partial_\beta\theta^a_\alpha)
\end{equation}
which is equivalent to:

\begin{equation}
\label{1.31.2}
\frac12 T_{\alpha\beta\gamma}dx^\alpha\wedge dx^\beta=g_{\gamma\rho}\theta^\rho_a d\theta^a=\eta_{bc}\theta^b_\gamma d\theta^c
\end{equation}
it follows that:

\begin{equation}
\label{1.31.3}
\frac12 T_{\alpha\beta\gamma}dx^\alpha\wedge dx^\beta\wedge dx^\gamma=\eta_{bc}d\theta^b\wedge\theta^c
\end{equation}
and therefore (\ref{3.2}) is equivalent to:

\begin{equation}
\label{1.32}
T_{[\alpha\beta\gamma]}=0
\end{equation}
which is also equivalent to:

\begin{equation}
\label{1.32.1}
T_{\alpha\beta\gamma}=K_{\alpha\beta\gamma}
\end{equation}
Either (\ref{1.32}) or (\ref{1.32.1})  are handy characterizations of Special Weitzenb\"{o}ck connections.

From (\ref{1.32.1}) we have:
 
\begin{equation}
\label{1.32.1.2}
g_{\gamma\rho}(\widetilde\Gamma^\rho_{\beta\alpha}-\widetilde\Gamma^\rho_{\alpha\beta})=
g_{\alpha\rho}(\widetilde\Gamma^\rho_{\beta\gamma}-\Gamma^\rho_{\beta\gamma})
\end{equation}
from where, clearing the Christoffel connection symbols, we obtain:

\begin{equation}
\label{1.32.2}
\Gamma^\sigma_{\beta\gamma}=\widetilde\Gamma^\sigma_{\beta\gamma}
+g_{\gamma\rho}g^{\alpha\sigma}(\widetilde\Gamma^\rho_{\alpha\beta}-\widetilde\Gamma^\rho_{\beta\alpha})
\end{equation}
or:

\begin{equation}
\label{1.32.3}
\Gamma_{\beta\gamma\alpha}=\widetilde\Gamma_{\beta\gamma\alpha}+\widetilde\Gamma_{\alpha\beta\gamma}-\widetilde\Gamma_{\beta\alpha\gamma}
\end{equation}
that can be considered also as a characterization of Special Weitzenb\"{o}ck connections.

The formula (\ref{1.32.2}) can not be inverted in general. To do it we need to demand a further restriction to the Weitzenb\"{o}ck connection. We shall say that the latter is Doubly special with Restricted covariance if systems of coordinates exist such that:

\begin{equation}
\label{1.39}
\widetilde\Gamma^\alpha_{\beta\gamma}=g^{\rho\alpha}g_{\sigma\beta}\widetilde\Gamma^\sigma_{\rho\gamma}
\end{equation}
or:

\begin{equation}
\label{1.39.1}
\widetilde\Gamma_{\alpha\gamma\beta}=\widetilde\Gamma_{\beta\gamma\alpha}
\end{equation}

Neither of these two equations are tensor equations. This is so because the l-h-s of (\ref{1.39}) is a connection but the r-h-s is not. If either of these equations is true in a given system of coordinates there will remain true only under the restricted group of coordinate transformations defined by the following conditions. 

\begin{equation}
\label{1.40}
g^{\rho\mu}\left(A^{\lambda^\prime}_\rho\partial_\gamma A^{\nu^\prime}_\mu
-A^{\nu^\prime}_\rho\partial_\gamma A^{\lambda^\prime}_\mu\right)=0
\end{equation}

Exchanging the indices $\alpha$ and $\beta$ in (\ref{1.32.3}) and adding we get:

\begin{equation}
\label{1.40.1}
\Gamma_{\beta\gamma\alpha}+\Gamma_{\alpha\gamma\beta}=\widetilde\Gamma_{\beta\gamma\alpha}+\widetilde\Gamma_{\alpha\gamma\beta}
\end{equation}
This equation combined with (\ref{1.39.1}) yields:

\begin{equation}
\label{1.41}
\widetilde\Gamma^\alpha_{\beta\gamma}=\frac12(\Gamma^\alpha_{\beta\gamma}+g_{\rho\beta}g^{\sigma\alpha}\Gamma^\rho_{\sigma\gamma})
\end{equation}
or, expanding the r-h-s:

\begin{equation}
\label{1.42}
\widetilde\Gamma_{\beta\gamma\alpha}=\frac12\partial_\gamma g_{\alpha\beta}
\end{equation}

If the orthonormal decomposition of a metric leads to an integrable Weit\-zenb\"{o}ck connection then there exists systems of coordinates such that:

\begin{equation}
\label{1.42.1}
\theta^a_\alpha=\chi_\alpha(x^\rho)\delta^a_\alpha, \quad g_{\alpha\beta}=\chi_\alpha^2\eta_{\alpha\beta}
\end{equation}
and a simple straightforward calculation proves that (\ref{1.39.1}) are verified and therefore this connection is doubly special with restricted covariance.


\section{Introduction of a second rank tensor}

At this point we assume that $n=4$. Let us consider  the dual tensors:

\begin{equation}
\label{1.34}
{*T}^{\alpha\rho\mu}=\frac12\eta^{\alpha\rho\gamma\delta}g^{\mu\nu} T_{\gamma\delta\nu},\ 
{*H}_{\alpha\rho\mu}=\frac12\eta_{\alpha\rho\gamma\delta}H^{\gamma\delta\nu}g_{\mu\nu}
\end{equation}
Using (\ref{1.17.0.3}) (\ref{1.17.0.4}), from (\ref{1.25}) and (\ref{1.30}) we get:

\begin{equation}
\label{1.35}
{\widetilde\nabla}_\alpha{*T}^{\alpha\rho\mu}=
\frac12\eta^{\rho\alpha\gamma\delta}{T}^\mu_{\sigma[\alpha}{T}^\sigma_{\gamma\delta]}, 
\ \ \widetilde\nabla_{[\alpha}{*H}_{\beta\rho]\mu}=\frac13\eta_{\alpha\beta\rho\sigma}(Y^\sigma_{\ \ \mu}-S^\sigma_\mu)
\end{equation}

We introduce the second rank tensor of type (1,1): 
\begin{equation}
\label{1.36}
U^\alpha_\beta=H^{\alpha\rho\mu} T_{\beta\rho\mu}+{*} T^{\alpha\rho\mu}{*H}_{\beta\rho\mu}
\end{equation}

The following two formulas follow from the preceding formulas (\ref{1.25}), (\ref{1.30}) and (\ref{1.35}):

\begin{equation}
\label{1.37.1}
\widetilde\nabla_\alpha(H^{\alpha\rho\mu} T_{\beta\rho\mu})=T_{\beta\rho\mu}(Y^\rho_{\ \ \nu}-S^\rho_\nu) g^{\mu\nu} +
\frac12 H^{\alpha\rho\mu}(\widetilde\nabla_\beta T_{\alpha\rho\mu}-3{T}^\nu_{\sigma[\alpha}{T}^\sigma_{\beta\rho]}g_{\mu\nu})
\end{equation}

\begin{equation}
\label{1.37.2}
\widetilde\nabla_\alpha({*T}^{\alpha\rho\mu} {*}H_{\beta\rho\mu})=
\frac12\eta^{\rho\alpha\gamma\delta}{T}^\mu_{\sigma[\alpha}{T}^\sigma_{\gamma\delta]}{*H}_{\beta\rho\mu}
+\frac12{*T}^{\alpha\rho\mu}(\nabla_\beta{*H}_{\alpha\rho\mu}
+\eta_{\alpha\beta\rho\sigma}(Y^\sigma_{\ \ \mu}-S^\sigma_\mu))
\end{equation}

And assuming now also that (\ref{1.32.1}) is satisfied, i.e. that the Weitzenb\"{o}ck connection is Special,  the tensor (\ref{1.30.1}) can be written as:

\begin{equation}
\label{1.37}
H_{\alpha\beta\delta}=T_{\alpha\beta\delta}-g_{\alpha\delta}T_\beta +g_{\beta\delta}T_\alpha,
\ \ T_\beta=-T^\rho_{\beta\rho}
\end{equation}
we have also:

\begin{equation}
\label{1.37.3}
H^{\alpha\rho\mu}\widetilde\nabla_\beta T_{\alpha\rho\mu}-T^{\alpha\rho\mu}\widetilde\nabla_\beta H_{\alpha\rho\mu}=0
\end{equation}

After some more elementary algebra we obtain:

\begin{equation}
\label{1.38}
\widetilde\nabla_\alpha U^\alpha_\beta=3H^{\sigma\rho\mu}{T}^\nu_{\alpha[\sigma}{T}^\alpha_{\beta\rho]}g_{\mu\nu}+2T_{\beta\sigma\mu}(Y^\sigma_{\ \ \nu} -S^\sigma_\nu) g^{\mu\nu}
\end{equation}
It is noteworthy that if Einstein's vacuum field equation are satisfied:

\begin{equation}
\label{1.38.1}
S_{\alpha\beta}=0
\end{equation}
then the divergence of the tensor $U^\alpha_\beta$ is cubic in the torsion while the tensor itself is quadratic.
The next section reviews the preceding ones in the frame-work of a linear approximation of General relativity that we considered in \cite{Bel}. It is in this simplified context that the relevance of the preceding tensor as a candidate to define some sort of energy-momentum is more explicit.


\section{The linear approximation}

In this section we assume that $\theta^a_\alpha$ and $\theta^\beta_b$ can be written as:

\begin{equation}
\label{2.1}
\theta^a_\alpha=\delta^a_\alpha+\frac12 f^a_\alpha, \  
\theta^\beta_b=\delta^\alpha_b+\frac12\bar f^\beta_b
\end{equation}
with:

\begin{equation}
\label{2.0}
\bar f^\beta_a\delta^a_\alpha+\delta^\beta_a f^a_\alpha=0
\end{equation}
$f^a_\alpha$ being four small quantities of, say, order $\epsilon$. Only the leading terms beyond the zero-order one will be kept in our calculations.

Consistently we shall consider only those coordinate transformations that are tangent to the identity transformation:

\begin{equation}
\label{2.3}
x^\rho=x^{\rho^{\prime}}+S^{\rho^\prime}(x^\prime),\ x^{\rho^{\prime}}=x^\rho-S^\rho(x)
\end{equation}
$S^\rho(x)$ being small quantities of order $\epsilon$, so that we shall have:

\begin{equation}
\label{2.4}
f^a_{\alpha^{\prime}}(x^{\prime})=f^a_\alpha(x)+\delta^a_\beta\partial_\alpha S^\beta(x)
\end{equation}

Again we shall consider only those frame transformations that are tangent to the identity transformation:

\begin{equation}
\label{2.7}
{f^\prime}^a_\alpha(x)=f^a_\alpha(x)+\delta^b_\alpha\Omega^a_b(x), 
\end{equation}
$\Omega^a_b$ being small quantities of order $\epsilon$ such that:

\begin{equation}
\label{2.6}
\eta_{cb}\Omega^c_a+\eta_{ac}\Omega^c_b=0
\end{equation}

The Riemannian metric defined in (\ref{1.17.0}) is now:

\begin{equation}
\label{2.2}
g_{\alpha\beta}=\eta_{\alpha\beta}+h_{\alpha\beta}, \ h_{\alpha\beta}=\eta_{ab}\delta^a_{(\alpha} f^b_{\beta)}, \ 
\end{equation}
and under a coordinate transformation we have:

\begin{equation}
\label{2.5}
h_{\alpha^{\prime}\beta^{\prime}}=h_{\alpha\beta}+\partial_{(\alpha}S_{\beta)}
\end{equation}

The Christoffel connection symbols are:

\begin{equation}
\label{2.9}
\Gamma_{\beta\gamma\rho}=\frac12(\partial_\beta h_{\gamma\rho}
+\partial_\gamma h_{\beta\rho}-\partial_\rho h_{\beta\gamma})
\end{equation}

\begin{equation}
\label{2.8}
\Gamma^\lambda_{\beta\gamma}=\Gamma_{\beta\gamma\rho}\eta^{\lambda\rho}
\end{equation}
and the Riemannian curvature is:

\begin{equation}
\label{2.11}
R^\alpha_{\beta\gamma\delta}=
\partial_\gamma\Gamma^\alpha_{\beta\delta}-\partial_\delta\Gamma^\alpha_{\beta\gamma} 
\end{equation}

The Weitzenb\"{o}ck connection is:

\begin{equation}
\label{2.12}
\widetilde\Gamma^\lambda_{\beta\gamma}=\frac12\delta^\lambda_a\partial_\gamma f^a_\beta, \ 
\widetilde\Gamma_{\beta\gamma\rho}=\frac12\partial_\gamma f^a_\beta\eta_{a\rho}
\end{equation}
or:

\begin{equation}
\label{2.12.1}
\widetilde\Gamma_{\beta\gamma\rho}=\frac12\partial_\gamma f_{\rho\beta}, \ f_{\rho\beta}=\eta_{a\rho}f^a_\beta
\end{equation}
It is convenient to use the symbol $f_{\rho\beta}$ but we have to keep in mind that this is not a second rank tensor. In fact under a coordinate transformation we have:

\begin{equation}
\label{2.12.2}
f_{\alpha^\prime\beta^\prime}(x^\prime)=f_{\alpha\beta}(x)+\partial_\beta S_\alpha(x)
\end{equation}
and under a frame transformation:

\begin{equation}
\label{2.12.3}
f^\prime_{\alpha^\prime\beta^\prime}(x)=f_{\alpha\beta}(x)+\Omega_{\alpha\beta}(x), \ 
\Omega_{\alpha\beta}+\Omega_{\beta\alpha}=0
\end{equation}

The  torsion of the Weitzenb\"{o}ck connection is:

\begin{equation}
\label{2.15}
T^\lambda_{\beta\gamma}=-\frac12\delta^\lambda_a(\partial_\gamma f^a_\beta-\partial_\beta f^a_\gamma)
\end{equation}
and the torsion identities are now:

\begin{equation}
\label{2.25}
\partial_{[\alpha}{T}^\rho_{\beta\gamma]}=0
\end{equation}

Using the general definition (\ref{1.17}) and the preceding relevant approximations we end up with this approximation of the contortion tensor:

\begin{equation}
\label{2.20}
K_{\alpha\beta\gamma}=\frac12\eta_{ab}(\delta^a_{[\alpha |} \partial_\gamma f^b_{|\beta]}
+\delta^a_{[\alpha}\partial_{\beta]} f^b_\gamma)
+\partial_{[\alpha} f^a_{\beta]}\delta^b_\gamma)
\end{equation}

Finally we list below the corresponding approximations of the general formulas derived in the preceding sections.  They are the following:

\begin{equation}
\label{2.26}
R^\alpha_{\beta\gamma\delta}=-\partial_\gamma K^\alpha_{\beta\delta}+\partial_\delta K^\alpha_{\beta\gamma}
\end{equation}

\begin{equation}
\label{2.27}
K_\beta=K^\sigma_{\beta\sigma}, \  K^\alpha=-\eta^{\rho\sigma}K^\alpha_{\rho\sigma}
\end{equation}

\begin{equation}
\label{2.28}
R_{\beta\delta}=-\partial_\alpha K^\alpha _{\beta\delta}+\partial_\delta K_\beta
\end{equation}

\begin{equation}
\label{2.29}
R=2\partial_\alpha K^\alpha
\end{equation}

\begin{equation}
\label{2.30}
S_{\beta\delta}=-\partial_\alpha H^\alpha_{\beta\delta}
\end{equation}
where:

\begin{equation}
\label{2.30.1}
H^\alpha_{\beta\delta}=K^\alpha_{\beta\delta}-\delta^\alpha_\delta K_\beta
+\eta_{\beta\delta} K^\alpha
\end{equation}
and:

\begin{equation}
\label{2.30.1.1}
\widetilde\nabla_\alpha U^\alpha_\beta=-2T_{\beta\sigma\mu}S^\sigma_\nu g^{\mu\nu}
\end{equation}
 
We prove now the main new result of this section. Namely, that there always exist, at the present approximation, Covariantly restricted Weitzenb\"{o}ck connections. 

Let us write:

\begin{equation}
\label{2.30.1.2}
T_{[\lambda\mu\nu]}=-\frac16\delta^{\alpha\beta\gamma}_{\lambda\mu\nu}\partial_\gamma f_{\alpha\beta}
\end{equation}
Generically this scalar will be different of zero, this meaning that the corresponding Weitzenb\"{o}ck connection is not Special, but under a frame transformation we shall have:

\begin{equation}
\label{2.30.2}
{T^\prime}_{[\lambda\mu\nu]}=-\frac{1}{6}\delta_{\lambda\mu\nu}^{\alpha\beta\gamma}\partial_\gamma (f_{[\alpha\beta]}+\Omega_{\alpha\beta})
\end{equation}
Therefore choosing:

\begin{equation}
\label{2.30.3}
\Omega_{\alpha\beta}=-f_{[\alpha\beta]}
\end{equation}
we select a Special Weitzenb\"{o}ck connection. Moreover this choice leads to:

\begin{equation}
\label{2.30.5}
\widetilde\Gamma^\prime_{\alpha\gamma\beta}-\widetilde\Gamma^\prime_{\beta\gamma\alpha}=0
\end{equation}
and therefore the connection is a covariantly restricted one. The restriction here comes from the fact that Eq. (\ref{2.30.3}) is not a tensor equation under the general coordinate transformations  that we considered in (\ref{2.3}). The covariance in this case is restricted to those transformations for which $S_\alpha$ is a gradient:

\begin{equation}
\label{2.30.6}
\partial_\alpha S_\beta-\partial_\beta S_\alpha=0
\end{equation}
and this is the basic assumption that we made in our paper \cite{Bel} to which the present one provides,we believe, a satisfactory geometrical justification.


\section*{Epilogue}

A bricklayer does not need a time-keeper as much as he needs a plumb line and a T-square, therefore we shall end this paper with some notes about the space geometry of space-time models.

Let us consider a time-like vector $\xi^\alpha$ and let $C$ be the congruence that it defines, $V_3$ being the corresponding quotient manifold. Using a system of adapted coordinates we have:

\begin{equation}
\label{E.1}
\xi^i=0,\quad i,j,k=1,2,3
\end{equation}
The line-element is then:

\begin{equation}
\label{E.2}
ds^2=-(-\xi dx^0+\xi_idx^i)^2+d\hat s^2,\ \ \xi=\sqrt{-g_{00}},\ \ \xi_i=\xi^{-1}g_{0i}
\end{equation}
and:

\begin{equation}
\label{E.3}
d\hat s^2=\hat g_{ij}dx^idx^j, \quad \hat g_{ij}=g_{ij}+\xi_i\xi_j
\end{equation}
Adapted coordinate transformations are those that leave invariant (\ref{E.1}). They are of the following form:

\begin{equation}
\label{E.4}
x^{0^\prime}=x^{0^\prime}(x), \ x^{i^\prime}=x^{i^\prime}(y) 
\end{equation}
where $y$ is a shorthand notation for $x^i$.

Quo-tensors are those objects, well-defined on $V_3$, that transform with an obvious rule generalizing the following one:

\begin{equation}
\label{E.5}
\hat g_{{i^\prime}{j^\prime}}(x^\prime)=\frac{\partial x^k}{\partial x^{i^\prime}}(y^\prime)
\frac{\partial x^l}{\partial x^{j^\prime}}(y^\prime)\hat g_{kl}(x(x^\prime))
\end{equation}

Two other important quo-tensors are the Newtonian-like field strength:

\begin{equation}
\label{E.6}
E_i=-(\partial_i\ln\xi+\hat\partial_0\xi_i)
\end{equation}
and the Coriolis rotation field:

\begin{equation}
\label{E.7}
B_{ij}=\hat\partial_{[i}\xi_{j]}-\xi_{[j}\partial_{i]}\ln\xi
\end{equation}
where $\hat\partial_\alpha$ are the symbols of the quo-differentiation operators:

\begin{equation}
\label{E.7.1}
\hat\partial_0=\xi^{-1}\partial_0, \ \ \hat\partial_i=\partial_i+\xi_i\hat\partial_0
\end{equation}
that are such that:

\begin{equation}
\label{E.8}
\hat\partial_{0^\prime}=\hat\partial_{0}, 
\ \ \hat\partial_{i^\prime}=\frac{\partial x^k}{\partial x^{i^\prime}}(y^\prime)\hat\partial_k
\end{equation}

It is also possible to consider quo-connections in this formalism. These are objects whose symbols transform as follows: 

\begin{equation}
\label{E.9}
\hat \Gamma^{p^\prime}_{{i^\prime}{j^\prime}}(x^\prime)=
\frac{\partial x^k}{\partial x^{i^\prime}}(y^\prime)
\frac{\partial x^l}{\partial x^{j^\prime}}(y^\prime)
\frac{\partial x^{p^\prime}}{\partial x^q}(y^\prime)
\hat \Gamma^q_{{kl}}(x^\prime)
+\frac{\partial x^{p^\prime}}{\partial x^{q}}(y(y^\prime))\frac{\partial^2 x^{q}}{\partial x^{i^\prime}\partial x^{j^\prime}}(y^\prime)
\end{equation}
under adapted coordinate transformations. Z'elmanov and Cattaneo introduced a Christoffel-like quo-connection whose symbols are those derived from the Christoffel symbols corresponding to the quotient metric (\ref{E.3}) with the substitution:

\begin{equation}
\label{E.10}
\partial_i \rightarrow \hat\partial_i
\end{equation}
so that:

\begin{equation}
\label{E.11}
\hat \Gamma^i_{jk}=\hat g^{is}(\hat\partial_j\hat g_{ks}+\hat\partial_k\hat g_{js}-\hat\partial_s\hat g_{jk})
\end{equation}

Similarly, we can consider a Weitzenb\"{o}ck-like quo-connection with symbols:

\begin{equation}
\label{E.12}
\bar \Gamma^i_{jk}=\theta^i_a\hat\partial_k\theta^a_j,\quad  a,b,..=1,2,3
\end{equation}
where $\theta^a_i$ are such that:

\begin{equation}
\label{E.13}
\hat g_{ij}=\delta_{ab}\theta^a_i\theta^b_j
\end{equation}

Defining now:

\begin{equation}
\label{E.14}
\hat d\theta^a=\frac12(\hat\partial_i\theta^a_j-\hat\partial_j\theta^a_i)dx^i\wedge dx^j
\end{equation}
and using obvious definitions of torsion $\bar T^i_{jk}$ and contortion $\bar K^i_{jk}$, the space-time definitions that we used before may be used now when dealing with the geometry of the quotient space $V_3$. We may thus say: (i) that the quo-connection (\ref{E.12}) is Special if:

\begin{equation}
\label{E.15}
\delta_{ab}\theta^a\wedge\hat d\theta^b=0;
\end{equation}
or equivalently if:

\begin{equation}
\label{E.18}
\bar T_{[ijk]}=0\Leftrightarrow \bar T_{ijk}=K_{ijk}
\end{equation}
(ii) that it is integrable if:

\begin{equation}
\label{E.16}
\theta^a\wedge \hat d\theta^a=0, \ \ a=1,2,3
\end{equation}
or (iii) that it is covariantly restricted if:

\begin{equation}
\label{E.17}
\bar \Gamma^i_{jk}=\frac12(\hat\Gamma^i_{jk}+\hat g^{is}\hat g_{rj}\hat\Gamma^r_{sk})
\end{equation}

There are two simple space geometries to consider: (i) that derived from an integrable congruence $C$:

\begin{equation}
\label{E.19}
B_{ij}=0
\end{equation}
and (ii) that derived from a rigid congruence in the sense of Born:

\begin{equation}
\label{E.20}
\partial_0 \hat g_{ij}=0
\end{equation}
In the first case there exist coordinates such that:   

\begin{equation}
\label{E.21}
\xi_i=0
\end{equation}
In the second case there exist orthonormal decompositions of $\hat g_{ij}$ such that: 

\begin{equation}
\label{E.22}
\partial_0\theta^a_i=0
\end{equation}
and in both cases we have:

\begin{equation}
\label{E.23}
\hat\partial_j\theta^a_i=\partial_j\theta^a_i
\end{equation}

More important. From Riemann's theorem about the diagonalization of three dimensional metrics we know that each of them can be derived from an orthogonal decomposition for which Eqs. (\ref{E.16}) are satisfied. Such is in particular the case when $C$ is a Born congruence. It appears then natural to generalize Born's definition of rigid motions requiring only the existence of orthonormal decompositions of the quotient metric such that Eqs. (\ref{E.16}) are satisfied. This definition is very similar and includes  that of \cite{Llosa}. 

\section*{Acknowledgements}

A. Molina has friendly contributed to improve a first version of this manuscript.


\end{document}